\documentclass[aps,prb,twocolumn,groupedaddress]{revtex4} 
\usepackage{amsbsy,amssymb,amsmath,bm}
\usepackage{graphicx,color,epsfig,rotate}

\begin{document}
\newcommand{\vp}{\varphi}
\newcommand{\be}{\begin{equation}}
\newcommand{\ee}{\end{equation}}
\newcommand{\bea}{\begin{eqnarray}}
\newcommand{\eea}{\end{eqnarray}}
\newcommand{\cP}{{\cal P}}
\newcommand{\cT}{{\cal T}}

\title{Compactons in ${\cal PT}$-symmetric generalized Korteweg-de Vries
Equations}

\author{Carl M. Bender${}^1$, Fred Cooper$^2$, Avinash Khare${}^3$, Bogdan
Mihaila${}^4$, and Avadh Saxena${}^5$}

\affiliation{${}^1$Department of Physics, Washington University, St. Louis, MO
63130, USA}
\email{cmb@wustl.edu}
\affiliation{${}^2$National Science Foundation, Division of Physics,
Arlington, VA 22230, USA and Santa Fe Institute, Santa Fe, NM 87501, USA}
\email{fcooper@nsf.gov}
\affiliation{${}^3$Institute of Physics, Bhubaneswar, Orissa 751005, India} 
\email{khare@iopb.res.in}
\affiliation{${}^4$Material Science and Technology Division, Los Alamos National
Laboratory, Los Alamos, NM 87545, USA}
\affiliation{${}^5$Theoretical Division and Center for Nonlinear Studies,
Los Alamos National Laboratory, Los Alamos, NM 87545, USA}
\email{bmihaila@lanl.gov}
\email{avadh@lanl.gov}

\date{\today}

\begin{abstract}
In an earlier paper Cooper, Shepard, and Sodano introduced a generalized KdV
equation that can exhibit the kinds of compacton solitary waves that were first
seen in equations studied by Rosenau and Hyman. This paper considers the $\cP
\cT$-symmetric extensions of the equations examined by Cooper, Shepard, and
Sodano. From the scaling properties of the $\cP\cT$-symmetric equations a 
general theorem relating the energy, momentum, and velocity of any solitary-wave
solution of the generalized KdV equation is derived, and it is shown that the
velocity of the solitons is determined by their amplitude, width, and momentum.
\end{abstract}


\maketitle

\section{Introduction}
\label{s1}
In a previous investigation Cooper, Shepard, and Sodano \cite{R1} introduced
and Khare and Cooper \cite{R2} studied further the first-order Lagrangian
\begin{equation}
L(l,p)=\int dx\left[\frac{\vp_x\vp_t}{2}+\frac{(\vp_x)^l}{l(l-1)}-\alpha
(\vp_x)^p(\vp_{xx})^2\right].
\label{e1}
\end{equation}
This Lagrangian is a modification of the original compacton equations of Rosenau
and Hyman \cite{R3}. (Unless otherwise specified, the range of $x$ integration
is over the entire real axis $-\infty<x<\infty$.) This Lagrangian gives rise to
a general class of KdV equations of the form
\begin{eqnarray}
u_t &+& u^{l-2}u_x+\alpha[2u^pu_{xxx}+4pu^{p-1}u_xu_{xx}\nonumber\\
&+& p(p-1)u^{p-2}(u_x)^3]=0,
\label{e2}
\end{eqnarray} 
where the solution $u(x,t)$ to the generalized KdV equation is defined by $u(x,t
)=\vp_x(x,t)$. For $0<p\leq2$ and $l=p+2$ these models admit compacton solutions
whose width is independent of the amplitude. For $p>2$ the derivatives of the
solution are not finite at the boundaries of the compacton where $u\to0$.
Cooper, Khare, and Saxena \cite{R4} analyzed the stability of the general
compacton solutions of this equation and showed that solutions are
stable provided that
\be
2<l<p+6.
\ee

There has been some recent interest in complex $\cP\cT$-symmetric extensions of
the ordinary KdV equation. Such extensions exist in the complex plane but also
lead to new PDEs that are entirely real. The first extension of the KdV equation
by Bender {\it et al.} \cite{R5} was
\be
u_t-iu(iu_x)^\epsilon+u_{xxx}=0,
\ee
which reduces to the usual KdV equation when $\epsilon=1$. This equation was
analyzed by Bender {\it et al.}\cite{R5} for $\epsilon=3$. This extension of
the KdV equation is not a Hamiltonian dynamical system at arbitrary $\epsilon$.
A more recent study by Fring \cite{R6} was based on a Hamiltonian formulation.
The Hamiltonian studied by Fring is related to a special case of the system
of generalized KdV equations examined here.

To find extensions of the generalized KdV equation that are invariant under the
joint operation of space reflection (parity) $\cP$ and time reversal $\cT$, we
make the following definitions: spatial reflection $\cP$ consists of making the
replacement $x\to-x$. Also, because $u$ is a velocity, under $\cP$ we replace
$u$ by $-u$. The effect of the time reversal operation $\cT$ is to change the
signs of $i$, $t$, and $u$: $i\to-i$, $t\to-t$, and $u\to-u$. Therefore, the
combination $iu_x$ is $\cP\cT$ even, so a $\cP\cT$-symmetric generalization of
the Lagrangian (\ref{e1}) is
\be
\label{e5}
L_{\cP\cT}=\int dx\left[\frac{\vp_x\vp_t}{2}+\frac{(\vp_x)^l}{l(l-1)}+\alpha(
\vp_x)^p(i\vp_{xx})^m\right].
\ee

For this Lagrangian we must find the correct $\cP\cT$-symmetric contour that
lies on the real axis when $m=2$. For $\cP\cT$ to be a good symmetry, branch
cuts must be taken along the positive imaginary axis in the
complex-$x$ plane. The Hamiltonian resulting from the above Lagrangian is
\be
H=\int dx\left[-\frac{u^l}{l(l-1)}-\alpha u^p(iu_x)^m\right].
\label{e6}
\ee
When $m$ is an even integer, a convenient choice for $\alpha$ that allows for
solitary-wave solutions and that gives a real equation for the generalized KdV
system is 
\be
-\alpha(m-1)i^m=1.
\label{e7}
\ee

For simplicity, we choose $\alpha$ as in (\ref{e7}) for most of this paper. The
$\cP\cT$ generalization of (\ref{e2}) has the same canonical structure as the
KdV equation. From Lagrange's equations or from Hamilton's equations for the
generalized KdV equations, we obtain the equations of motion for $u(x,t)$:
\be
\frac{\partial u}{\partial t}=\frac{\partial}{\partial x}\frac{\delta H}{\delta
u}=\{u,H\},
\ee
where the Poisson bracket structure is \cite{R7}
\be
\{u(x),u(y)\}=\partial_x\delta(x-y).
\ee
The resulting equation becomes
\bea
\label{eq}
&& 0=u_t+u_xu^{l-2}+u^{p-2}u_x^{m-3}\left[(m-2)mu^2 u_{xx}^2\right.\nonumber\\
&& \left.+2mpuu_{xx}u_x^2+mu^2u_{xxx}u_x+(p-1)pu_x^4\right].
\label{e10}
\eea
 
This system of equations has three obvious conservation laws: conservation of
mass $M$, momentum $P$, and energy $E$, where the energy is the value of the
Hamiltonian (\ref{e6}) and
\be
\label{mom}
M=\int dx\,u(x,t),\quad P=\int dx\,\frac{1}{2}u^2(x,t).
\ee
The case $m=2$ leads to the well known compacton solutions. Fring \cite{R6}
studied a Hamiltonian similar to the subclass of this $\cP\cT$-symmetric class
of Hamiltonians corresponding to $l=3$ and $p=0$, but with slightly different
coefficients for the two terms in the Hamiltonian.

This paper is organized as follows: Section~\ref{s2} considers the scaling
properties of the nonlinear wave equation (\ref{e10}) and discusses the energy
and momentum of solitary waves, and Sec.~\ref{s3} continues the discussion of
these conserved quantities. Traveling-wave solutions are discussed in
Sec.~\ref{s4} and the conserved quantities for these solutions are examined in
Sec.~\ref{s5}. Some special cases are described in Sec.~\ref{s6}. The question
of stability is addressed in Secs.~\ref{s7} and \ref{s8}.

\section{scaling properties}
\label{s2}
Let us examine the scaling properties of (\ref{e10}). We require that solutions
transform into solutions under the scaling
\be
x\to\lambda x,\quad t\to\lambda^\eta t,\quad u\to\lambda^\beta u, 
\ee
and we find that
\be
\beta-\eta=(l-1)\beta-1=\beta(p+m-1)-m-1.
\ee
Solving for $\beta$ we obtain
\be
\beta=\frac{m}{p+m-l}.
\ee
We also find that
\be
1-\eta=\beta(l-2).
\ee

Suppose that we have a traveling solitary wave of the form $f(x-ct)$. Then $c$
scales as $x/t$ or as $\lambda^{1-\eta}$. Therefore, $c$ scales as
\be
c\propto\lambda^{\beta(l-2)}.
\ee
In terms of the velocity, $x$ scales as $\lambda$ and 
\be
\lambda\propto c^{i_1},\quad i_1=\frac{p+m-l}{m(l-2)}.
\ee
From the equation for $i_1$ we see that the width of the solitary wave does not
depend on the velocity when
\be
l=p+m.
\ee
This is a generalization of the result for $m=2$. The conserved momentum scales
like $u^2x\propto\lambda^{2\beta+1}$, so
\be
P\propto c^{i_2},\quad i_2=\frac{3m-l+p}{m(l-2)}.
\ee

Finally, the conserved energy scales as $u^l x$, so
\be 
E\propto\lambda^{\beta l+1}\propto c^{(ml+p+m-l))/(p+m-l)}.
\ee
Eliminating $c$ in favor of $P$ in this formula, we get
\be
E\propto P^{-r},
\ee
where 
\be
r=-\frac{lm+p+m-l}{p+3m-l}.
\label{e22}
\ee
This reduces to our previous results for the case $m=2$.

We show below how to make these scaling laws more precise and how to determine
the constants of proportionality that relate the conserved quantities. 

\section{Another Way to Relate Energy and Momentum of Solitary Waves}
\label{s3}
In Ref.~[4] a general theorem was derived that relates the energy, momentum, and
velocity of solitary waves of the generic form $u(x,t)=AZ[\beta(x-q(t))]$ for
$m=2$. Here, we generalize this result to arbitrary $m$ and we use the results
obtained by studying the scaling properties of the alternative action $\Phi$ to
justify the previous derivation. 

We start from the action
\be
\Gamma=\int dt\,L,
\ee
where $L$ is given in (\ref{e5}). We now assume that the exact solitary-wave
solution has the generic form
\be
\label{e99}
\phi_{x}\equiv u=AZ[\beta(x-ct)].
\ee
Using this form, it is easy to calculate the value of the Hamiltonian (\ref{e6})
for the solitary wave (\ref{e99}):
\be
H=-C_{1}(l)\frac{A^l}{\beta l(l-1)}+A^{p+m} \beta^{m-1} C_2(p,m),
\ee
where
\be
C_1(l)=\int dz\,Z^{l}(z),\quad C_2(p,m)=\int dz\,[Z'(z)]^mZ^{p}.
\label{e102}
\ee
Since $H$ and momentum $P$ are conserved, we can rewrite the parameter $A$ in
terms of $P$:
\be
P=\frac{1}{2}\int dx\,u^2=\frac{A^2}{2\beta}C_{5},\quad C_{5}=\int dz\,Z^2(z).
\label{e103}
\ee
Replacing $A$ by $P$, we rewrite the Hamiltonian $H$ as
\be
H=-C_3(l)P^{\frac{l}{2}}\beta^{\frac{l-2}{2}}+C_4(p,m)P^{\frac{p+m}{2}}
\beta^{\frac{p+3m-2}{2}},
\ee
where
\bea
C_3(l)&=&\frac{C_{1}(l)}{l(l-1)}\sqrt{2/C_{5}},\nonumber\\
C_4(p,m)&=&\alpha C_2(p,m)\left(2/C_5\right)^{(p+m)/2}.
\eea
At this point, we note that 
the exact solutions have the property that they are
the functions of the parameter $\beta$ that minimize the Hamiltonian with
respect to $\beta$ when the momentum $P$ is fixed. Using $\partial H/\partial
\beta=0$, we obtain
\be
\beta=P^{\frac{p+m-l}{l-p-3m}}\left[\frac{C_4(p,m)(p+3m-2)}{C_3(l)
(l-2)}\right]^{\frac{2}{l-p-3m}}.
\ee
This leads to
\be
H=f(l,p,m)P^{-r},
\ee
where $r$ is given by (\ref{e22}) and 
\bea
&& f(l,p,m)=C_{3}(l)\frac{p+m-l}{p+3m-2}\nonumber\\
&& \times\,\left[\frac{C_{4}(p,m)(p+3m-2)}{C_{3}(l)(l-2)}\right]^{\frac{l-2}{
l-p-3m}}.
\label{e107}
\eea
Hamilton's equation $\dot{q}=\partial H/\partial P$ yields the relationship
\be
-\dot{q}=c=rH/P.
\ee
From this analysis it is again easy to show that the momentum $P$, amplitude
$A$, and width parameter $\beta$ functionally depend on the velocity $c$ (note
that $c=-\dot{q}$):
\be
\label{e109}
P\propto c^{\frac{p+3m-l}{m(l-2)}},\quad A\propto c^{\frac{1}{l-2}},\quad\beta
\propto c^{\frac{l-p-m}{m(l-2)}}.
\ee
Here, the proportionality constants depend on $C_{i}(l,p,m)$ ($i=1,2,3,4,5$)
defined above, and once an exact solution is obtained, these constants can be
calculated easily. 

We make three observations: (i) When $l=p+m$, the width parameter $\beta$ is
independent of the velocity $c$ and momentum $P$ and hence of the amplitude $A$
of the solitary wave. (ii) The $c$ dependence of the amplitude $A$ depends
solely on the parameter $l$, and it is independent of the parameters $p$ and
$m$. (iii) The stability problem when $l=p+2$ was studied by Dey and Khare
\cite{R11} for the case $m=2$ using the results of Karpman\cite{R10}. In
Sec.~VII we extend their results to arbitrary even integer $m$.

\section{Traveling-wave solutions} 
\label{s4}
We begin with the wave equation (\ref{e10}) which can be reexpressed as
\bea
&& u_t+u^{l-2}u_x-\frac{p}{m-1}\left[u^{p-1}(u_x)^m\right]_x\nonumber\\
&& +\frac{m}{m-1}\left(u^p u_x^{m-1}\right)_{xx}=0,
\label{eqlmp2}
\eea
and assume that 
\be
u(x,t)=f(x-ct)\equiv f(y).
\ee
Then,
\bea	
&& cf'=f^{l-2}f'+\nonumber\\
&& \frac{1}{m-1}\left(p\left[f^{p-1}(f')^m\right]'-m\left[f^p(f')^{m-1}\right]''
\right),
\label{e25}
\eea
and integrating once we obtain
\be
cf=\frac{f^{l-1}}{l-1}+mf^p(f')^{m-2}f''+p(f')^mf^{p-1}+K_1,
\label{e26}
\ee
where $f'\equiv df/dy$. 

For compact solutions $K_1=0$. Setting the integration constant $K_1$ to zero,
multiplying this equation by $f$, and integrating over $y$, we obtain
\be\label{e27}
cI_2=\frac{1}{l-1}I_l-\frac{p+m}{m-1}J_{m,p},
\ee
where
\be\label{e28}
I_n=\int_{-\infty}^\infty dy\,f^n(y),\quad J_{m,p}=\int_{-\infty}^\infty dy\,
(f')^m f^p(y).
\ee 

For noncompact solutions if $K_1\neq0$, we also have a term that includes $I_1$,
which is the conserved mass. We multiply (\ref{e26}) by $f'$ and integrate again
with respect to $y$ to get the following nonlinear differential equation for the
traveling waves:
\be
\frac{c}{2}f^{2}-\frac{f^l}{l(l-1)}-(f')^mf^p=K_1f+K_2.
\ee
We see that $K_2$ must also be zero for solutions $f$ that are compact. Now, if
we set $K_1=0$ and $K_2=0$ and integrate with respect to $y$, we obtain
\be
J_{m,p}=\frac{c}{2}I_2-\frac{1}{l(l-1)}I_l.
\label{e30}
\ee

From ({\ref{e27}) and (\ref{e30}) we can solve for $J_{m,p}$ and $I_l$ in terms
of $I_2$. We obtain
\be
J_{m,p}=\frac{(l-2)(m-1)}{2[p+m+(m-1)l]}cI_2
\label{e31}
\ee
and
\be\label{e32}
I_l=\frac{l(l-1)(p+3m-2)c}{2[p+m(l+1)-l]}I_2.
\ee

Notice that when $l\to2$, $c=1$ for consistency. This is related to the fact
that the equation for $u(x,t)$ becomes a linear equation with propagation
velocity $c=1$. The energy of the solitary wave is given by
\be\label{e33}
H=\frac{1}{m-1}J_{m,p}-\frac{1}{l(l-1)}I_l
\ee
and the momentum $P=I_2/2$. From Eqs. (\ref{e31}) and (\ref{e33}) we deduce 
that the energy, momentum, and
velocity of the solitary wave are related by
\be
H=Pc/r,
\ee
where $r$ is given in (\ref{e22}). 
 
\subsection{Weak solutions}
We are interested in compacton solutions that are a combination of a compact
function $f(x)$ confined to a region (initially $-x_0<x<x_0$ and zero
elsewhere). At the boundaries $\pm x_0$ the function $f(x)$ is assumed to be
continuous but higher derivatives most likely are not. For there to be a weak
solution we require that the jump in
\be
\frac{c}{2}f^2-\frac{f^l}{l(l-1)}-(f')^mf^p-K_1f-K_2
\label{e35}
\ee
be zero when we cross from $x_0-\epsilon$ to $x_0+\epsilon$. Since $f(x_0)$ is
assumed to vanish, the requirement for a weak solution is
\be
Disc[(f')^m(x)f^p(x)]_{x_0}=0,
\ee
where $Disc$ is the discontinuity across the boundary $x_0$. This is
always satisfied if there is no infinite jump in the derivative of the function.
 
We are concerned mostly with the cases where the integration constants are set
equal to zero. We can then rewrite (\ref{e35}) as
\be
\frac{c}{2}f^{2-p}-\frac{f^{l-p}}{l(l-1)}=(f')^m.
\label{e37}
\ee
Notice that for $m=2$ we recover
\be
\frac{c}{2}f^{2-p}-\frac{f^{l-p}}{l(l-1)}=(f')^2,
\label{e38}
\ee
which was studied previously. The solitary wave for (\ref{e38}) is obtained by
joining the positive and negative solutions of the square root of (\ref{e38}),
and for $m=2n$ one again obtains a real solution by joining the positive and
negative real parts of the solutions of the $m$th root of (\ref{e37})
appropriately shifted so that the maximum is at the origin $y=0$.

If we are looking for compactons, then the finiteness of the derivative when
$f\to0$ requires that 
\be
p\leq 2,\quad p\leq l.
\label{e39}
\ee
The case $m=2$ gives both compactons and real equations. For $m=2$, when
\be
l=p+2,
\ee
the width of the compacton is independent of the velocity. In our previous
discussion of scaling we found that when $l=p+m$, the width of the compacton is
indeed independent of the velocity. 

\subsection{Compacton solutions when $m$ is an even integer}

Compacton solutions are constructed by patching a compact portion of a periodic
solution that is zero at both ends to a solution that vanishes outside the
compact region to give the weak solution described above. Let us look at
the generalizations of the compacton equation when we go from $m=2$ to $m=4$.
Consider the case when $p=1$ and let $l=3$ and $l=4$. For $p=1$, $l=3$ the
equation for the solitary wave is
\be
\frac{c}{2}f-\frac{1}{6}f^2=(f')^m.
\ee

For the positive branch of the solution, we get
\be
x-ct=\int_0^f\frac{du}{\left(\frac{c}{2}u-\frac{1}{6}u^2\right)^{1/m}}.
\ee
Performing the integral, we obtain
\be
x-ct=2^{\frac{1}{m}}3^{\frac{m-1}{m}}c^{\frac{m-2}{m}} B_{\frac{f}{3
c}}\left(\textstyle{\frac{m-1}{m}},\textstyle{\frac{m-1}{m}}\right),
\ee
where $B_{n} (x,y)$ is the incomplete beta function while $B(x,y)$ (see
below) is the complete beta function.

For $m=2$ this simplifies to 
\be
x-ct=2\sqrt{6}\sin^{-1}\left[\sqrt{f/(3c)}\right],
\ee
which leads to the previous compacton result
\be
f=3c\sin^2\left[\textstyle{\frac{1}{2\sqrt{6}}}(x-ct)\right].
\ee
For $m=4$ we get for the positive real fourth root
\be
x-ct=2^{1/4}\,3^{3/4}\,\sqrt{c}\left[B_{\frac{f}{3c}}\left(\textstyle{\frac{
3}{4}},\textstyle{\frac{3}{4}}\right)\right].
\ee

In Fig.~\ref{f1} we plot $B_{\frac{f}{3 c}}\left(\frac{3}{4},\frac{3}{4}\right)
-B\left(\frac{3}{4},\frac{3}{4}\right)$ and its mirror image as a function of
$f/(3c)$. Here, $y=(x-ct)2^{-1/4}3^{-3/4}c^{-1/2}$.
\begin{figure}
\begin{center}
\epsfig{figure=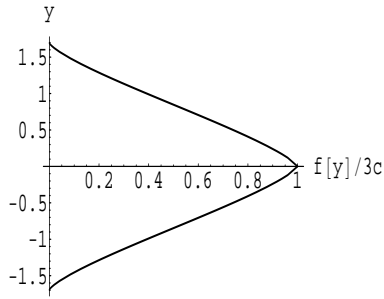,width=2.0 in,height=2.0 in}
\caption{$f/(3c)$ versus $y$ for $m=4$, $l=3$, and $p=1$.}
\label{f1}
\end{center}
\end{figure}

Now consider the case $p=1$ and $l=4$, where the solitary-wave equation becomes
\be
\frac{c}{2}f-\frac{1}{12}f^3=(f')^m.
\ee
For the positive branch of the solution, we get
\be
x-ct=\int_0^f du \big [\frac{c}{2}u-\frac{1}{12}u^3 \big ]^{-1/m},
\ee
or
\be
x-ct=(3/2)^{\frac{m-1}{2m}}c^{\frac{m-3}{2m}}B_{\frac{f^2}{6c}}\left(
\textstyle{\frac{m-1}{2m}},\textstyle{\frac{m-1}{m}}\right).
\ee
By adding the positive-real- and negative-real-root solutions (for $m$ an even
integer) we get the complete compacton profile. The compacton vanishes
elsewhere.

For $m=2$ this leads to 
\be
x-ct=[3/(2c)]^{1/4}B_{\frac{f^2}{6c}}\left(\textstyle{\frac{1}{4}},
\textstyle{\frac{1}{2}}\right).
\ee
When m=2, the equation for the compacton can in fact be directly solved in 
terms of elliptic function. In particular, consider the equation at $m=2$ 
\be
\frac{c}{2}f-\frac{1}{12}f^3=(f')^2.
\ee
Assuming a solution of the form
\be
f=A\,{\rm cn}^2(\beta y,k^2=1/2),
\ee
we find that 
\be
\beta=c^{1/4}(96)^{-1/4},\quad A=\sqrt{6c}, 
\ee
where ${\rm cn}(x,k)$ is the Jacobi elliptic function with modulus $k$. 

\subsection{Hyperelliptic compactons}
Consider next the generalization of the hyperelliptic compactons discussed in
Ref.~[4]. For this purpose we assume that we can parametrize the
solutions by
\be
f=AZ^a[\beta(x-ct)],
\ee
and we demand that 
\be
(Z')^m=1-Z^{2\tau}.
\label{e55} 
\ee
This immediately leads to the relations
\be\label{e56}
a=\frac{m}{m+p-2},\quad\tau=\frac{m(l-2)}{2(m+p-2)}.
\ee
We also find that
\be\label{e57} 
A^ma^m\beta^m=\frac{c}{2}A^{2-p}=\frac{A^{l-p}}{l(l-1)}.
\ee
This leads to 
\be
A=[cl(l-1)/2]^{1/(l-2)}
\label{e58}
\ee
and
\be
\beta=\frac{1}{a[l(l-1)]^{1/m}}[cl(l-1)/2]^{(l-p-m)/(m(l-2))}.
\label{e59}
\ee
Note that this ansatz gives the correct scaling behavior of the amplitude
parameter $A$ and the width parameter $\beta$ for the velocity $c$. 

The solution to the differential equation (\ref{e55}) has $m$ branches
corresponding to the various values of $e^{2i\pi n/m}$ when $m$ is an integer
and $n=1,\,2,\,\ldots,\,m$. For even integer $m$, the positive root can be
integrated to give
\bea
y &=& \int_0^Z dx\,\left(1- x^{2\tau}\right)^{-1/m}\nonumber\\
&=&Z\,_2F_1\left(\frac{1}{m},\frac{1}{2\tau};1+\frac{1}{2\tau};Z^{2\tau}\right),
\eea
where ${}_2F_1$ denotes the hypergeometric function. For even $m$ we get the
full solution for the compacton by adding the positive-real-root and the
negative-real-root solutions to get the complete compacton profile. The
compacton vanishes elsewhere. 

\section{Conserved quantities} 
\label{s5}
For solutions satisfying (\ref{e55}) it is possible to determine explicitly the
conserved quantities in terms of the velocity $c$ of the wave and the
parameters $l$, $p$, and $m$ of the differential equation. We have already found
that the parameters $A$ and $\beta$ are given by (\ref{e58}) and (\ref{e59}).

We use the generic integral
\be
\int_0^1 dz\,z^\alpha(1-z^{2\tau})^\beta
=\frac{\Gamma\left(\frac{\alpha+1}{2\tau}
\right)\Gamma(\beta+1)}{2\tau\Gamma\left(\frac{\alpha+1}{2\tau}+\beta+1\right)}.
\ee
For the mass we get
\bea 
M&=&\int dx A Z^a (\beta x)=\frac{A}{\beta}\int dZ(dZ/dy)^{-1} Z^a\nonumber\\
&=& 2\frac{A}{\beta}\int_0^1 dZ\,Z^a(1- Z^{2 \tau})^{-1/m}\nonumber\\
&=& \frac{A\Gamma\left(\frac{\text{m}-1}{\text{m}}\right)\Gamma\left(\frac{a+1}
{2\tau}\right)}{\beta\tau\Gamma\left(1-\frac{1}{m}+\frac{1+a}{2\tau}\right)}.
\eea
Here, we have used the fact that the total area under the solitary wave is twice
the area coming from the positive root. (See Fig.~\ref{f1}.) For the momentum we
get
\bea 
P&=&\frac{1}{2}\int dx\,A^2Z^{2a}(\beta x)=\frac{A^2}{\beta}\int dZ\,(dZ/dy)^{
-1}Z^{2a}\nonumber\\
&=&\frac{A^2}{\beta}\int_0^1 dZ\,Z^{2a}(1-Z^{2\tau})^{-1/m}\nonumber\\
&=&\frac{A^2\Gamma\left(\frac{m-1}{m}\right)\Gamma\left(\frac{2a +1}{2\tau}
\right)}{2\beta\tau\Gamma\left(1-\frac{1}{m}+\frac{1+2a}{2\tau}\right)},
\eea
where $a=m/(m+p-2)$ and the energy is $E=cP/r$. 

\section{Special Cases}
\label{s6}
There are two types of special cases. The first occurs when $\tau$, $l$, and $p$
are integers. From Eq. (\ref{e56}) it follows that for $m=4$
\be
l=2+\tau(p+2)/2.
\ee
We consider here the set $p=\{1,2\}$. For $p=1$, $\tau=2n$ and $l=3n+2$. For $p=
2$, we find instead that $\tau$, $l=2n+2$, $n=1,\,2,\,\ldots$. For $m=6$
\be
l=2+\tau(p+1)/3.
\ee
Thus, $\tau=3n$, $l=2+n(p+1)$ yields simple integer solutions for integer $n$,
$p=1,2$. 

The other interesting case arises when the width of the solitary wave is
independent of the velocity. This occurs when
\be
l=p+m.
\ee
We study the cases for which $\{p,l\}=\{1,1+m\}$ and $\{p,l\}=\{2,2+m\}$. When
$l=p+m$, $\tau$ is given by
\be
\tau=m/2
\ee
and here we also have $a=m/(l-2)$.

\subsection{Case $m=4$}
For this case
\be
\tau=\frac{2l-4}{p+2},\quad a=\frac{4}{p+2}.
\ee
For $\tau=2$, $l=p+4$ and the width is independent of velocity. The relevant
function to invert is
\be
y=\int_0^Z \frac{dx}{(1- x^4 )^{1/4}}=Z\,_2F_1\left(\textstyle{\frac{1}{4}},
\textstyle{\frac{1}{4}};1+\textstyle{\frac{1}{2\tau}};Z^{4}\right).
\ee 

For a compacton centered about the origin $y=0$, the two halves of the
compacton are given by 
\be
y_\pm=\pm f_1(Z)\mp f_1(Z=1),
\ee
where
\be
f_1(x)=x\, _2F_1\left(\textstyle{\frac{1}{4}},\textstyle{\frac{1}{4}};
\textstyle{\frac{5}{4}};x^4\right)
\ee
and 
\be
f_1(Z=1)=\Gamma\left(\textstyle{\frac{3}{4}}\right)\Gamma\left(\textstyle{
\frac{5}{4}}\right)=\textstyle{\frac{1}{4}}\pi\sqrt{2}= 1.11072...\,.
\ee
The result for $Z[y]$ is shown in Fig. 2.
\begin{figure}
\begin{center}
\epsfig{figure=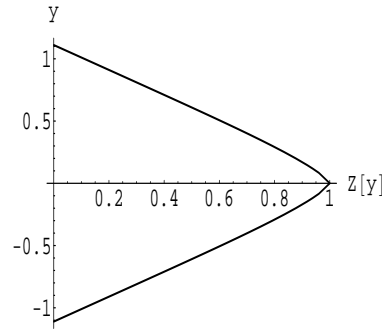,width=2.0 in,height=2.0 in}
\caption{$Z[y]$ for $m=4$, $l=p+4$ $\tau=2$.}
\label{fig2}
\end{center}
\end{figure}

For the case $p=1$, $l=5$, we get $A=(c/2)^{1/3}$ and the solution goes as $Z^{4
/3}$. For the case $p=2$, $l=6$, we get $A=(c/2)^{1/4}$ and the solution is
linear in $Z$.

When $\tau=3$, there is another special case with integer values: $p=2$; $l=8$,
$A=1$. Then, the relevant function to invert is
\be
y=\int_0^Z dx\left(1-x^6\right)^{-1/4}=Z\, _2F_1\left(\textstyle{\frac{1}{6}},
\textstyle{\frac{1}{4}};\textstyle{\frac{7}{6}};Z^6\right).
\ee 
Since
\be
_2F_1\left(\textstyle{\frac{1}{6}},\textstyle{\frac{1}{4}};\textstyle{\frac{7}
{6}};1\right)=\Gamma\left(\textstyle{\frac{3}{4}}\right)\Gamma\left(\textstyle{
\frac{7}{6}}\right)/\Gamma\left(\textstyle{\frac{11}{12}}\right),
\ee
we again use both the positive and negative solutions to make up the entire 
function $Z(y)$. The case $p=0$ does not correspond to compacton solutions and
$p>2$ does not allow for a finite derivative when the amplitude of the solitary
wave becomes zero. 

\subsection{Case $m=6$}
For integer $\tau=3n$ we discuss the $\tau=3$ case. Here, $l=p+3$, and there are
two possibilities: Let $p=1$, $l=4$. Then,
\be
A\propto c^{1/2},\quad a=6/5.
\ee
For $p=2$, $l=5$,
\be
A\propto c^{1/3},\quad a=1. 
\ee

The form of the function $Z[y]$ is now similar to the previous case:
\be
f_2(Z)=\int_0^Z dx\left(1-x^6\right)^{-1/6}=Z\, _2F_1\left(\textstyle{\frac{1}
{6}},\textstyle{\frac{1}{6}};\textstyle{\frac{7}{6}};Z^6\right).
\ee
We obtain the two halves of the function $Z(y)$ by inverting
\be
y_\pm=\pm f_2(Z)\mp f_2(Z=1),
\ee
where
\be
f_2(Z=1)=\Gamma\left(\textstyle{\frac{5}{6}}\right)\Gamma\left(\textstyle{
\frac{7}{6}}\right)=\textstyle{\frac{\pi}{3}}=1.0472...\,.
\ee

\section{Alternative Generating Function and Stability}
\label{s7}
Solitary waves of the form $f(y)=f(x-ct)$ can be derived by considering the
following function:
\bea
\Phi[f(y),f'(y)]&=&\int dx\left(H[f,f']+P[f]c\right)\nonumber\\
&\equiv&\int dx\,\varphi[f,f'].
\eea
Notice that $\varphi$ is the negative of the Lagrangian density. That is, the
original equation for the solitary wave can be written as 
\be
\partial_x\frac{\delta\Phi}{\delta f}=0.
\ee
The once-integrated equation ({\ref{e26}}) (with no integration constants) is
obtained from
\be
\frac{\delta\Phi}{\delta f}=0,
\ee
or equivalently from the Euler-Lagrange equation
\be
\frac{\partial \varphi}{\partial f}=\frac{d}{dx}\left(\frac{\partial\varphi}
{\partial f'}\right).
\ee

We have explicitly
\be
\Phi=\int dy\left[-\frac{f^l}{l(l-1)}+f^p(f')^m+\frac{1}{2}cf^2\right].
\ee
The first variation after an integration by parts can be written as
\bea
\delta\Phi&=&\int dy\Bigg[-\frac{f^{l-1}}{l-1}+cf-pf^{p-1}(f')^m\nonumber\\
&&\quad -mf^p(f')^{m-2}f''\Bigg]\delta f.
\eea
The second variation which is important for the linear stability analysis can be
written as
\be
\delta^2\Phi=\int dy\,\delta f~L~\delta f,
\ee
where $L$ is the operator
\bea
L&=&c-f^{l-2}-p(p-1)f^{p-2}(f')^m-mpf^{p-1}(f')^{m-2}f''\nonumber\\
&&\left(-mpf^{p-1}(f')^{m-1}-m(m-2)f^p(f')^{m-3}f''\right)\frac{d}{dy}
\nonumber\\
&&-mf^p(f')^{m-2}\frac{d^2}{dy^2}.
\label{eq:L}
\eea
When $m=2$, this reduces to the result given in Dey and Khare \cite{R11}.

One can write $\Phi$ in terms of $I_l$ and $J_{m,p}$:
\be
\Phi[f]=\frac{1}{m-1}J_{m,p}-\frac{1}{l(l-1)}I_l+\frac{c}{2}I_2.
\ee
Following Derrick \cite{R8}, we consider the scale transformation $x\to\lambda
x$. Under this transformation
\be
I_l[f(\lambda y)]=\frac{1}{\lambda}I_l,\quad J_{m,p}[f(\lambda y)]=\lambda^{m-1}
J_{m,p},
\ee
so that
\be
\Phi[f(\lambda y)]=\frac{1}{m-1}\lambda^{m-1}J_{m,p}-\frac{I_l}{\lambda l(l-1)}
+\frac{c}{2\lambda}I_2.
\ee
If we assume that taking the derivative of $\Phi$ with respect to $\lambda$ and
setting $\lambda=1$ gives a solution, we get
\be
\frac{d\Phi(\lambda)}{d\lambda}\Big|_{\lambda=1}=0=J_{m,p}+\frac{I_l}{l(l-1)}-
\frac{c}{2}I_2.
\label{e87}
\ee
This is precisely (\ref{e30}), the equation of motion integrated over space
that we found earlier. [Derrick looked to see if the second derivative of
(\ref{e87}) became negative which would indicate that the solitary wave was
unstable.] If we calculate the second derivative, we obtain
\bea
&& \frac{d^2\Phi(\lambda)}{d\lambda^2}=cI_2\frac{l-2}{2l(l-1)[l(-1+m)+m+p]}
\left[m^2l^2\right.\nonumber\\
&&\!\!\left.-ml^2-\left(m^2-7m-2p+4\right)l+6m+2p-4\right].
\eea

This does not factor to give a simple criterion for stability. However, another
choice leads to a simple stability criterion. Suppose we instead make the
scaling
\be
f(y)\to\lambda^\rho f(\lambda y).
\ee
This again leads to the equations of motion plus a boundary term because
\bea
\frac{d\Phi}{d\lambda}\Big|_{\lambda=1}&=&\int dy\left[\frac{\partial\varphi}{
\partial f}-\frac{d}{dx}\frac{\partial\varphi}{\partial f'}\right]\left(\rho f+x
f'\right)\nonumber\\
&+&\left[\frac{\partial\varphi}{\partial f'}\right](\rho f+xf'')\Big|_{y_{\rm
 min}}^{y_{\rm max}}=0. 
\eea
Assuming that the boundary term vanishes at the edges of the compacton, we
recover the equation of motion
\bea
\Phi[\lambda^{\rho}f(\lambda y)] &=& \frac{1}{m-1}\lambda^{m-1+\rho(m+p)}
J_{m,p}\nonumber\\
&&-\frac{I_l}{l(l-1)}\lambda^{l\rho-1}+\frac{c}{2}\lambda^{2\rho-1}I_2.
\eea
The condition for a minimum is 
\bea
&&\frac{d\Phi(\lambda)}{d\lambda}\Big|_{\lambda=1}=0=\frac{(l\rho-1)I_l}{l(l-1)}
\nonumber\\
&&-\frac{m-1+\rho(m+p)}{m-1}J_{m,p}-\frac{c}{2}(2 \rho-1)I_2.
\eea

The particular case $\rho=1/2$ is special in that the conserved momentum $P$ is
invariant under this transformation; that is,
\be
P[\lambda^{1/2}f(\lambda y)]=P[f(y)].
\ee
If we choose $\rho=1/2$, when we vary $\Phi$, we are varying the Hamiltonian
with the constraint that $P$ is held fixed. This is exactly what happens in a
trial variational calculation where the parameter $\lambda$, now thought of as a
variational parameter, is a constraint variable to be eliminated and determined
in terms of the momentum $P$, which is the dynamical variable of the reduced
Hamiltonian system. (This particular variation was first considered by
Kuznetsov \cite{R9} and was then elaborated on by Karpman\cite{R10} and Dey
and Khare \cite{R11}.)

For $\rho=1/2$ we obtain
\be
J_{mp}=\frac{(l-2)(m-1)}{l(l-1)(3m+p-2)}I_l,
\ee
which is precisely the relation obtained by using (\ref{e31}) and (\ref{e32}).
For arbitrary $\rho$ we get a linear combination of (\ref{e26}) and (\ref{e30}).
For arbitrary $\rho$ the second derivative does not factor into a simple form
that allows one to say when it changes sign. However, for $\rho=1/2$ the answer
does factor and the second derivative yields 
\be
\Phi''(\lambda)\Big|_{\lambda=1}=\frac{Pc(l-2)(3m+p-l)(3m+p-2)}{4l(-1+m)+m+p}.
\ee

We also learn from the conditions in (\ref{e39}) that a weak solution that is
compact can exist if $p\leq 2,~p\leq l$. This leads to the statement that
solitary waves will be unstable under this type of deformation when
\be
l>p+3m. 
\ee
More general scale transformations involving two parameters, such as 
\be
f(y)\to\mu^{1/2}f(\lambda y),
\ee
have been discussed by Karpmann \cite{R10} and Dey and Khare \cite{R11}.
 
\subsection{Linear Stability}
In this section we extend the analysis of Karpman \cite{R10} to our generalized
KdV equation. To study linear stability we assume that we can write 
\be
u(x,t)=f(y)+v(x,t),\quad|v|\ll1,\quad(u,v)=0,
\ee
where 
\be
(f,g)=\int_{-\infty}^\infty dx\,f^\ast g.
\ee
We change variables to $y=x-vt$ and $T=t$, so that we parametrize an arbitrary
addition to $f(y)$ as $v(y,T)$. The linearized equation of motion for $v(y,t)$
is then
\be
\frac{\partial}{\partial T}v(y,T)=\frac{\partial}{\partial y}\left[Lv(y,T)
\right],
\ee
where $L$ is given by (\ref{eq:L}).

The equation (\ref{e25}) for the solitary wave can be written as
\be
L\partial_y f(y)=0.
\ee
Thus, $f'(y)$ is a zero eigenfunction of $L$ corresponding to the translation
invariance of the solution (that is, the Goldstone mode). 

Now, if we take the derivative of the first integral of the solitary wave
equation (\ref{e26}) with respect to the velocity $c$, we obtain
\be
L\frac{\partial f}{\partial c}=-f.
\label{eq:cderiv}
\ee
Inverting, we get
\be
\frac{\partial f}{\partial c}=-L^{-1}f.
\label{useful}
\ee
This result will be useful later. We also have
\be
(f,f')=0
\ee
from integrating by parts. If we now consider 
\be
v(y,T)=e^{-i\omega T}\psi(y)+e^{i\omega^\star T}\psi^\star(y),
\ee
then $\psi$ satisfies
\be
\omega\psi(y)=i\partial_y L\psi(y).
\label{eq:stab}
\ee
When $m$ is an even integer, $L$ is a Hermitian operator. In that case there is
a theorem that all the $\omega$ are real if one of the two operators on the
right side of (\ref{eq:stab}) is positive definite \cite{R14}. (The more general
case will be studied elsewhere.) A sufficient condition for real eigenvalues is
\be
(\psi,L\psi)>0,
\ee
where $\psi$ is orthogonal to $f$ and $f'$. This condition is exactly the same
as requiring that the second variation of $\Phi=H+Pc$ be a minimum at $f$. If
the solitary-wave solution is a minimum of $\Phi$, then the solution is linearly
stable. 

Our objective is to find the extremal value of $(\psi,L\psi)$ and to find the 
criterion that guarantees that it is positive. To find the extrema one
solves the constrained variation condition
\be
\delta[(\psi,L\psi)-\Lambda(\psi,\psi)-C(\psi,f)]=0,
\ee
so that 
\be
(L-\Lambda)\psi=Cf.
\ee
Using $(\psi,f)=0$, we find that
\be
(\psi,L\psi)=\Lambda,\quad(\psi,\psi)=1.
\ee

One solves this equation by expanding $\psi$ and $f$ in a series of
eigenfunctions of the operator $L$. Letting $L\phi_n=\lambda_n\phi_n$ and
assuming the ordering $\lambda_n>\lambda_m$ if $n>m$, we find that $\lambda_1=0$
and $\phi_1= f'$. Letting
\be 
f=\sum_{n\neq1}b_n\phi_n,
\ee
we find that
\be
\psi=C\sum_{n\neq1}\frac{b_n}{\lambda_n-\Lambda}\phi_n.
\ee
From $(\psi,f)=0$ we obtain
\be
r(\Lambda)\equiv\sum_{n\neq1}\frac{|b_n|^2}{\lambda_n-\Lambda}\phi_n=0.
\ee
We are interested in the lowest solution $\Lambda_{min}$ that solves this
equation because this gives the minimum of $(\psi,L\psi)$ that we seek. Assuming
along with Karpman \cite{R10} as well as Dey and Khare \cite{R11} that $L$ has
only one negative eigenstate, one then deduces that $(\psi,L\psi)>0$ is
satisfied if $r(0)<0$. However, when $\Lambda=0$ we get 
\be
L\psi_{\Lambda=0}=Cf.
\ee
Thus, from (\ref{eq:cderiv}) we find that
\be
\psi_{\Lambda=0}=\frac{\partial f}{\partial c}
\ee
and
\be
r(0)=-\bigg(\frac{\partial f}{\partial c},f\bigg)=-\frac{\partial P}
{\partial c}.
\ee
This criterion, namely that
\be
\frac{\partial P}{\partial c}>0
\label{Pstability}
\ee
for stability, gives exactly the same sufficient result for stability as all the
other criteria used.

Since for all of our solutions $P\propto c^{(p+3m-l)/[(l-2)m]}$, it immediately
follows that these solutions are linearly stable provided that
\be
2<l<p+3m.
\label{e111}
\ee

\subsection{Lyapunov Stability} 
Lyapunov stability uses sharp estimates and has been used by Weinstein
\cite{R15} and Karpman et al. \cite{R16}. Here we want to show that the
compacton solution is a minimum of the Hamiltonian for fixed momentum $P$. We
show this using Holder's Inequality \cite{R17} and follow the arguments of
Weinstein \cite{R15}, Kuznetzov \cite{R9}, Karpmann \cite{R10}, and Dey and
Khare \cite{R11}. We do this by first showing that the Hamiltonian for fixed
momentum $P$ is bounded below and then that the compacton solution satisfies the
condition that it is a particular lower bound. We can write the Hamiltonian $H$
in terms of $I_l$ and $J_{m,p}$ as
\be
\label{e122}
H[f]=\frac{1}{m-1}J_{m,p}-\frac{1}{l(l-1)}I_l, 
\ee
where 
\be
I_2=2P.
\ee
Now consider that
\be
I_l\leq\left({\rm max}~f^{l-2}\right)\int dy\,f^2. 
\ee
We want to bound $I_l$ by a function of $J_{m,p}$ and the conserved momentum
$P$. To do this it is convenient to write
\be
f^{l-2}=\left[f^a\right]^{(l-2)/a}.
\ee

Then, writing
\be
f^a=\int dy\,df^a/dy=a\int dy\,f^{a-1-k}\frac{df}{dy}f^k,
\ee
we use the Holder inequality \cite{R17} to show that 
\bea
f^a &\leq& a\left(\int dy\,|f^{(a-1-k)m}(f')^m|\right)^{1/m}\nonumber\\
&&\times\left(\int dy\,|f^{kj}(y)|\right)^{1/j}.
\eea
We can dispense with the absolute values when $m$ is an even integer. Thus by
choosing
\be
kj=2;\quad p=m(a-1-k),
\ee
we can relate the second term of (\ref{e122}) to $P$ and $ J_{m,p}$.

Moreover, to relate the bound to the energy of the solitary wave (rather than
having a general lower bound that depends on the choice of $a$), one must
further choose
\be
ma=p+3m-2,
\ee
from which we find that 
\be
j=\frac{m}{m-1}.
\ee
Doing this we obtain
%%% (since the power of $P$ is $(m-1)(l-2)/(ma)+1/2$)
\be
H\geq{\rm min}\Bigg|_{J_{m,p}}\frac{J_{m,p}-\alpha_{p,l,m}J_{m,p}^{(l-2)/
(p+3m-2)}(2P)^\gamma}{m-1}, 
\ee
where 
\be
\gamma=\frac{(l-2)(m-1)+p+3m-2}{p+3m-2},
\ee
and
\be
\alpha_{p,l,m}=\frac{1}{l(l-1)}\left(\frac{p+3m-2}{m}\right)^{m(l-2)/(p+3m-2)}.
\ee
Minimizing with respect to $J_{m,p}$ for fixed momentum $P$, we obtain
\be
H_{min}=\frac{l-p-3m}{(m-1)(l-2)}J_{m,p}.
\ee

For the solitary wave which obeys the generalized KdV equation we get
\be
J_{m,p}=\frac{(l-2)(m-1)}{p+m+(m-1)l}P_{sol}c.
\ee
Thus, the solitary wave is a minimum of the Hamiltonian and satisfies
\be
E_{sol}=E_{min}=P_{sol}c/r,
\ee
where $r$ is as given by (\ref{e22}). Thus, as long as $2<l<p+3m$, one has
stable solitary waves. 

\section{Variational stability of solutions}
\label{s8}

Suppose that we have found an exact solution of the form $AZ\{\beta[x-q(t)]\}$.
Then one can find sufficient conditions for instability of this type of solution
by seeing if the solution is a minimum rather than a maximum of the Hamiltonian
as a function of $\beta$ with the conserved momentum $P$ held fixed.
 
We know that the exact solutions are stationary for fixed $P$ under variations
in $\beta$. That is,
\be
\frac{\partial H}{\partial\beta}=0.
\ee
We can write the Hamiltonian in the generic form ($P$ fixed)
\be
H=-C_1\beta^a+C_2\beta^b,
\ee
where the constants depend on $P$ and the parameters that define the Lagrangian
and $a$ and $b$ also depend on the parameters in the Lagrangian.

The stationarity condition is
\be
\frac{\partial H}{\partial\beta}=0=\frac{1}{\beta}\left[-C_1a\beta^a
+C_2b\beta^b\right],
\ee
from which we infer that
\be
C_1a\beta^a=C_2b\beta^b.
\ee
The edge of stability of these solutions is given by
\be 
\frac{\partial^2 H}{\partial\beta^2}=0=-C_1a(a-1)\beta^a+C_2b(b-1)\beta^b.
\ee
At the minimum this leads to the condition that
\be
a=(l-2)/2,\quad b=(p+3m-2).
\ee
Thus, the critical case is
\be 
l=p+3m,
\ee
which agrees with (\ref{e111}). We expect that the solutions we have found are
stable as long as 
\be
l<p+3m.
\ee

\subsection{Approximate variational solutions}

To study stability it is useful to have approximate solutions that are close to
the exact solutions to see if they relax to the exact solutions or become
unstable. For this purpose it is useful to study the {\it post-Gaussian} trial
functions
\be
f_V(x-ct)\equiv g(x-ct)=A\exp\left[-|\beta(x-ct)|^{2n}\right],
\ee
where $A$, $\beta$, and $n$ are continuous variational parameters chosen to
minimize the action. These trial wave functions have earlier been successfully
used \cite{R13} to approximate various solitary waves in both KdV systems and
NLSE applications.

The advantage of these trial functions is that the action as well as all of the
conserved quantities can be explicitly evaluated using the formula
\be
\int_0^{\infty}dx\,x^a e^{-b|x|^{2n}}=\frac{1/2n}b^{-\frac{a+1}{2n}}\Gamma
\left(\textstyle{\frac{a+1}{2n}}\right).
\ee

In terms of our previous notation we have
\be
Z(z)=A\exp\left(-|z|^{2n}\right).
\ee
Thus, using the trial function, we get
\bea
C_1(l)&=&\int dz\,Z^l(z)=2\int_0^\infty dz\,e^{-l|z|^{2n}}\nonumber\\
&=& l^{-\frac{1}{2 n}}\Gamma\left(1+\frac{1}{2n}\right).
\eea
We also have 
\bea
&& C_2(p,m)=\int dz\,[Z'(z)]^m Z^p\nonumber\\
&& \,=2\int_0^\infty dz\,(-2)^m e^{-mx^{2n}-p}n^mx^{m(2n-1)}\nonumber\\
&& \,=\frac{1}{n}(-2n)^m(m+p)^{\frac{m-2nm-1}{2n}}\Gamma\left(m-\textstyle{
\frac{m-1}{2n}}\right).
\eea
These expressions depend on the variational parameter $n$, which determines the
shape of the solution; $n=1/2$ gives a {\it peakon} shape and $n=1$ is the usual
Gaussian. From these we can determine the quantities
\be
C_4(p,m)=K(p,m,n)\Gamma\left(1+\textstyle{\frac{1}{2n}}\right)^{-\frac{m+p}{2}}
\Gamma\left(m-\textstyle{\frac{m-1}{2n}}\right),
\ee
where
\be
K(p,m,n)=\frac{(-n)^{m-1}2^{\frac{6nm+m+2np+p}{4n}}(m+p)^{\frac{m-2nm-1}{2n}}}
{m-1}
\ee
and 
\be
C_3(l) = \frac{2^{\frac{2 n l+l}{4 n}} l^{-1-\frac{1}{2 n}}\Gamma
\left(1+\frac{1}{2 n}\right)^{1-\frac{l}{2}}}{l-1}.
\ee

For the trial function we get 
\be
A=\sqrt{2\beta P/C_5}
\ee
and
\be
\beta=P^{\frac{p+m-l}{l-p-3m}}\left[\frac{C_4(p,m)(p+3m-2)}{C_3(l)(l-2)}
\right]^{\frac{2}{l-p-3m}}.
\ee
To determine the best trial function in this class we must also minimize the
Hamiltonian with respect to the parameter $n$. As in our discussion of
conservation laws, we get
\be
H=f(l,p,m,n)P^{-r},
\ee
where $r$ is given in (\ref{e22}) and $f(l,p,m,n)$ is given in (\ref{e107}).

For solutions that are compact and cover half of the period of a positive
periodic function, an alternative choice for a variational trial function is
\be
u_2(x)=A[\cos(\beta x)]^\gamma,
\ee
where $\beta$ and $\gamma$ are the variational parameters. For integer $p,m$,
and $l$ it is again possible to obtain an explicit expression for $H[\beta,
\gamma]$. One can perform the minimization with respect to $\beta$ explicitly.
Determining the global minimum in the parameter $\gamma$ must be done
numerically.

\subsection{Case $p=1$, $l=3$, $m$}
First, consider $m=2$, where the exact solitary wave solution is 
\be
f(y)=3c\sin^2\left(\frac{y}{2\sqrt{6}}+\frac{\pi}{2}\right).
\ee
For simplicity we normalize our functions by choosing $P=1$. Then, this solution
is
\be
f(Z)=2^{5/4}3^{-3/4}\pi^{-1/2}\cos^2\left(\frac{z}{2\sqrt{6}}\right).
\ee
This belongs to the class of variational solutions of the second type, and we
would have obtained this exact answer from our variational minimization
procedure. For the post-Gaussian trial functions, the lowest-energy variational
solution having $P=1$ is found to be
\be
u(x)=0.583578\,\exp\left(-0.0314705\,x^{2.308}\right).
\ee
Fig.~3 compares the exact and variational function for $P=1$. Note that apart
from the region where the compacton goes to zero the agreement is excellent.

\begin{figure}
\begin{center}
\epsfig{figure=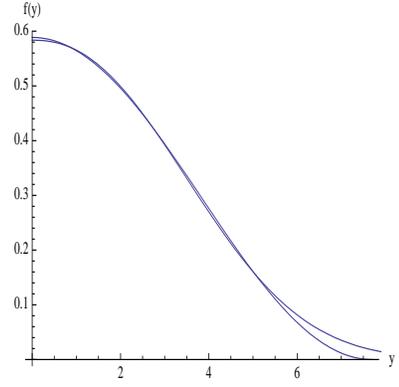, width=2.0 in, height=2.0 in}
\caption{$f(y)$ versus $y$ for $m=2$, $l=3$, $p=1$. This graph compares the
exact versus variational solitary waves for $P=1$.}
\label{fig1}
\end{center}
\end{figure}

For $m=4$ the value of $n$ that minimizes the first trial function is $n=
0.920655$. Again, normalizing to $P=1$ (which yields $c\approx 1/3$), we
find that the best function in this class is:
\be
u(x)=0.995936\exp\left(-0.396108\,x^{1.84131}\right).
\ee
If we use the second type of trial function we find that the values of $\beta$
and $\gamma$ that give a global minimum in the reduced space are
\be
\beta=0.342787,\quad\gamma=5.67846,
\ee
which leads to $A=0.97067$ for $P=1$. Thus, the best trial function in the 
second class is given by 
\be
u_2(x)=0.97067\,[\cos(0.342787 x)]^{5.67846}.
\ee

In Fig.~4 we compare the two variational approximations and note that apart
from the fact that the $u(x)$ is not compact, the agreement is quite good. Both
solutions are global minima of the respective reduced Hamiltonians, which depend
on two parameters. 
\begin{figure}
\begin{center}
\epsfig{figure=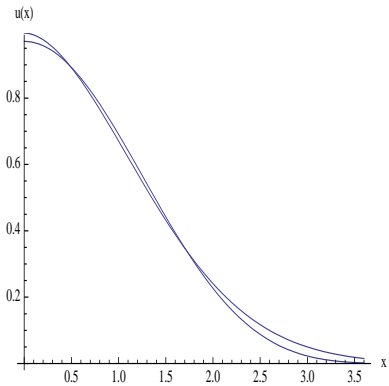,width=2.0 in, height=2.0 in}
\caption{$u(x)$ and $u_2(x)$ versus $x$ for $m=4$, $l=3$, $p=1$.}
\label{fig4}
\end{center}
\end{figure}

\newpage
To compare our implicit exact result with our variational approximations, we
change variables to $y=x/2^{1/4}3^{3/4}c^{1/2}$ and redisplay the approximate
solitary waves shown in Fig.~5 along with the exact result of Fig.~1. 
\begin{figure}
\begin{center}
\epsfig{figure=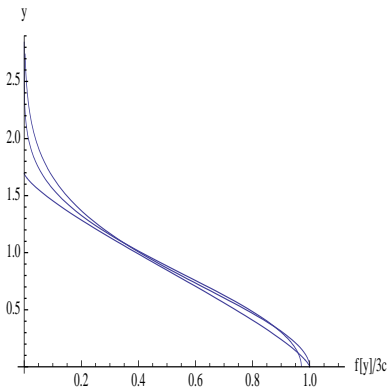,width=2.0 in,height=2.0 in}
\caption{$f(y)/3c$ versus $y$ for $m=4$, $l=3$, $p=1$. Exact answer versus two
variational solitary wave solutions.}
\label{fig5}
\end{center}
\end{figure}


\begin{thebibliography}{99}
\bibitem{R1} F.~Cooper, H.~Shepard and P.~Sodano, Phys.~Rev.~E {\bf 48}, 4027
(1993).
\bibitem{R2} A.~Khare and F.~Cooper, Phys.~Rev.~E {\bf 48}, 4843 (1993). 
\bibitem{R3} P.~Rosenau and J.~M.~Hyman, Phys.~Rev.~Lett.~{\bf 70}, 564 (1993). 
\bibitem{R4} F.~Cooper, A.~Khare, and A.~Saxena, Complexity {\bf 11}, 30 (2006).
\bibitem{R5} C.~M.~Bender, D.~C.~Brody, J.-H.~Chen, and E.~Furlan,
J.~Phys.~A: Math.~Theor.~{\bf 40}, F153 (2007).
\bibitem{R6} A.~Fring, J.~Phys.~A: Math.~Theor.~{\bf 40}, 4215 (2007).
\bibitem{R7} A.~Das, {\it Integrable Models} (World Scientific, Singapore,
1989).
\bibitem{R11} B.~Dey and A.~Khare, Phys.~Rev.~E {\bf 58}, R2741 (1998).
\bibitem{R10} V.~I.~Karpman, Phys.~Lett.~A {\bf 210}, 77 (1996).
\bibitem{R8} G.~H.~Derrick, J.~Math.~Phys.~{\bf 5}, 1252 (1964).
\bibitem{R9} E.~A.~Kuznetsov, Phys.~Lett.~A {\bf 101}, 314 (1984).
%\bibitem{R12} B.~Dey, Phys.~Rev.~E {\bf 57}, 4733 (1998).
\bibitem{R14} I. M. Gel'fand, {\it Lectures on linear algebra} (Interscience,
New York, 1961), Sec.~15.
\bibitem{R15} M.~I.~Weinstein, Commun.~Math.~Phys.~{\bf 87}, 567 (1983).
\bibitem{R16} V.~I.~Karpman, Phys.~Lett.~{\bf A215}, 254 (1996) and references
therein.
\bibitem{R17} O.~Holder, Nachr.~Ges.~Wiss.~Gottingen {\bf 38} (1889). See also
G.~H.~Hardy, J.~E.~Littlewood, and G. Polya, {\it Inequalities} (Cambridge
University Press, Cambridge, 1934).
\bibitem{R13} F.~Cooper, H.~Shepard, C.~Lucheroni, and P.~Sodano, Physica D
{\bf 68}, 344 (1993); F.~Cooper, C.~Lucheroni, H.~Shepard, and P.~Sodano,
Phys.~Lett.~A {\bf 173}, 33 (1993).

\end{thebibliography}
\end{document}